\documentclass[journal,twocolumn]{IEEEtran}
\usepackage{graphicx}
\usepackage[caption=false]{subfig}
\usepackage{lipsum}
\usepackage{afterpage}
\usepackage{amsmath}
\usepackage{amssymb}
\usepackage{cite}
\PassOptionsToPackage{hyphens}{url}
\usepackage[hidelinks]{hyperref}
\usepackage{booktabs}
\usepackage{algorithm}
\usepackage{algpseudocode}
\usepackage{caption}
\usepackage{float}
\usepackage{stfloats}
\usepackage{siunitx}
\usepackage{multirow} 
\usepackage{makecell} 
\usepackage{array} 

\begin{document}

\title{Latency-Aware Deep Learning Benchmark for Real-Time Cyber-Physical Attack and Fault Classification in Inverter-Dominated Power Grids}

\author{
\IEEEauthorblockN{Emad Abukhousa, Saman Zonouz, and A.P. Sakis Meliopoulos}
\IEEEauthorblockA{\textit{ \\School of Electrical and Computer Engineering, Georgia Institute of Technology, Atlanta, GA, USA \\}
\{emadak, szonouz6, sakis.m\}@gatech.edu}
}

\maketitle

\begin{abstract}
This work introduces a latency-aware benchmarking framework for evaluating deep learning models in power system anomaly detection using high-fidelity, time-domain signals generated from an industry-grade electromagnetic transient simulator. Eight neural network architectures—ranging from MLPs to Transformers—were systematically evaluated on streaming datasets representing both physical faults and cyber-attacks in inverter-dominated networks. All models successfully classified two representative multi-event sequences in real time with sub-cycle response times ($<$15 ms). However, although classification decisions occurred within one cycle, the end-to-end inference latency consistently exceeded three cycles (50–90 ms). These results highlight a critical gap between algorithmic capability and protection-grade deployment, pointing to the need for further optimization and hardware acceleration. The findings establish a reproducible benchmark for sub-cycle anomaly detection and provide guidance for transitioning machine learning methods from research prototypes to real-world protection applications.
\end{abstract}


\section{Introduction}

Modern power systems are undergoing a profound transformation, driven by the deep integration of inverter-based resources (IBRs) and the deployment of sophisticated cyber-physical infrastructure. This evolution enhances operational efficiency and sustainability but concurrently introduces complex, intertwined failure modes. IBRs, such as solar photovoltaics and battery energy storage systems, exhibit fault current characteristics—including limited magnitudes and high-frequency harmonics—that challenge legacy protection schemes designed for the predictable behavior of synchronous machines \cite{Reno2021IBRPart1}. Simultaneously, the digitalization of substations through standards like IEC 61850 expands the cyber-attack surface, creating vulnerabilities to threats that can manipulate physical processes \cite{Zonouz2012SCPSE}. This convergence of physical and cyber complexities necessitates a new paradigm for anomaly management, one that can rapidly and accurately distinguish between diverse events to ensure grid resilience.

\vspace{10pt}
The operational timeline for power system protection is exceptionally demanding, often requiring fault isolation within one to two cycles (16–34 ms) to prevent equipment damage and cascading failures. This need for speed is equally critical for cybersecurity, where the timely detection of a malicious data stream can avert widespread disruption. The availability of high-fidelity data from Merging Units (MUs), which stream time-synchronized Sampled Values (SV) at rates like 4.8 kHz, provides an unprecedented opportunity to detect anomalies at their inception \cite{Meliopoulos2023rCSP}.

The application of DL to power system anomalies shows considerable promise, successfully detecting and classifying physical faults \cite{Thomas2023TIM_CNNTransformerFaults} and cyber-attacks \cite{Almalaq2022DMLSmartGrid} with high offline accuracy. Research has demonstrated its feasibility on high-frequency EMT-grade data \cite{Khaw2021TSG} and validated models in HIL environments \cite{Roy2023_DLRelay_Microgrid}. However, a critical gap persists: most studies evaluate models on static datasets, prioritizing accuracy over end-to-end inference latency in a realistic streaming context. This oversight is a major barrier to deployment, as a slow model is operationally ineffective regardless of its accuracy.

This deficiency is highlighted in recent reviews, which identify the lack of efficiency metrics and latency-aware reporting as a systemic weakness in the field \cite{Moriano2025_AAD_SLR, MishraSingh2025-DL-PSP-Review, Hasan2024_EnergyReports_ML_SG_CPS}. They conclude that claims of "real-time" applicability are rarely substantiated with the rigorous timing analysis required for protection-grade systems. The absence of a comparative benchmark for sub-cycle latency leaves a key question unanswered: which models can actually meet modern power systems' stringent timing requirements?

This paper directly addresses this gap by introducing a comprehensive, latency-aware benchmarking framework to evaluate the real-time performance of deep learning models for power system protection. The primary contributions are: (1) a systematic, head-to-head comparison of eight distinct DL architectures on their ability to detect and classify 17 different physical faults and cyber-attacks using high-fidelity (4.8 kHz) streaming data; (2) the implementation of a practical real-time inference pipeline featuring a one-cycle moving average filter and a confidence-based abstention policy to enhance decision stability; and (3) the first quantitative, comparative evidence of sub-cycle detection and classification latencies for this diverse suite of models. By providing concrete data on the trade-offs between accuracy, speed, and complexity, this work offers actionable insights for designing and deploying the next generation of intelligent, resilient, and fast-acting protection systems.

\section{Methodology}

\subsection{High-Fidelity Dataset Generation}

The dataset was generated using the industry-grade simulator WinIGS \cite{winigs2025} , which models grid dynamics, renewable integration, and anomaly injections with microsecond precision. Data acquisition emulated IEC~61850 Sampled Value (SV) streams, with time-domain waveforms extracted from COMTRADE files. These records capture the digitized outputs of current and potential transformers (CTs/PTs) connected to two Merging Units (MUs), preserving substation-grade measurement fidelity.

\begin{figure}[!t]
\centering
\includegraphics[width=\columnwidth]{
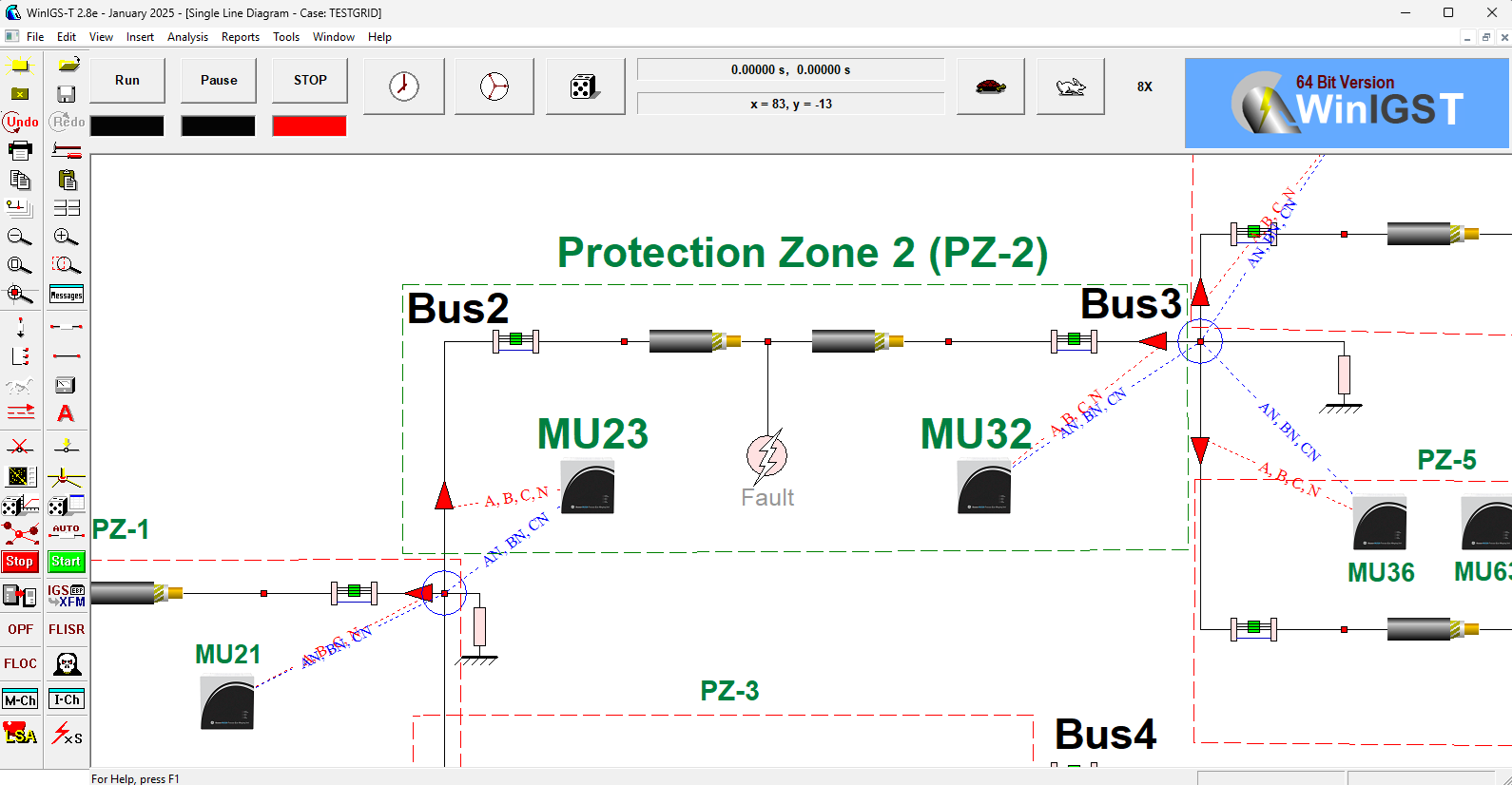}
\caption{Testbed configuration highlighting Protection Zone~2 (PZ-2) with MUs MU23 and MU32 and representative fault location.}
\label{fig:testbed}
\end{figure}
\vspace{5pt}
The testbed (Fig.~\ref{fig:testbed}) represents an \emph{inverter-rich microgrid} with distributed energy resources (DERs), including a 50~kVA photovoltaic inverter and a 30~kVA battery energy storage system (BESS), alongside step-down transformers, distribution lines, and diverse loads. This composition reflects modern DER-integrated grids where inverter control dynamics dominate transient behavior.

\vspace{5pt}
For this study, Protection Zone~2 was instrumented with two MUs (MU23, MU32) providing 14 synchronized features. Simulations produced 105,768 samples over 22~s at 4.8~kHz, yielding a temporal resolution of 208~$\mu$s. Each MU supplied seven CT/PT-derived signals: three-phase currents ($I_a$, $I_b$, $I_c$), neutral current ($I_n$), and three-phase voltages ($V_{an}$, $V_{bn}$, $V_{cn}$). Together, these time-domain waveforms enable detailed analysis of inverter-driven dynamics, harmonic distortion, and fault transients.

\subsection{Deep Learning Model Training}

Eight deep learning architectures, representing distinct paradigms in neural network design, were evaluated: Multi-Layer Perceptron (MLP), 1D Convolutional Neural Network (1D-CNN), Long Short-Term Memory (LSTM), CNN-LSTM hybrid, Transformer, Temporal Convolutional Network (TCN), ResNet1D, and Spatio-Temporal Graph Convolutional Network (ST-GCN).
\vspace{5pt}

Preprocessing included handling missing values, standardizing current and voltage measurements using a scaler fitted exclusively on the training split, and encoding anomaly class labels into integer indices. This ensured consistent scaling and prevented information leakage. To capture temporal dependencies, feature sequences of 50 samples were generated with stride 1 using a sliding window approach.

The dataset was partitioned via stratified block-based splitting into training (70\%), validation (10\%), and test (20\%) sets, while class weights were applied to mitigate imbalance. All models were trained for 40 epochs using the Adam optimizer and sparse categorical cross-entropy loss with class weighting. Callbacks for early stopping and learning rate reduction were employed to improve convergence and reduce overfitting.

\begin{table}[!t]
\small
\caption{Complete anomaly taxonomy}
\label{tab:anomaly_taxonomy}
\begin{center}
\begin{tabular}{clc}
\toprule
Class & Anomaly Type \\
\midrule
0 & Normal Operation \\
1-3 & SLG Faults (A-N, B-N, C-N) \\
4,8 & CT Ratio Attacks (MU23, MU32) \\
5-7 & LL Faults (AB, AC, BC) \\
9-11 & DLG Faults (AB-N, AC-N, BC-N) \\
12-13 & PT Ratio Attacks (MU23, MU32) \\
14-15 & Three-Phase Faults (ABC, ABC-N) \\
16-17 & GPS Spoofing Attacks (MU23, MU32) \\
\bottomrule
\end{tabular}
\end{center}
\end{table}

\subsection{Real-Time Streaming Evaluation}

The streaming pipeline operates directly on raw time-domain samples from IEC~61850 Merging Units (SV format), ensuring transient waveforms are preserved. At each sample $i$, the model produces a probability vector $\mathbf{P}(i) \in [0,1]^K$, which is smoothed using a one-cycle centered moving average:

\begin{equation}
P_{\text{avg}}[i,k] = \frac{1}{N_{\mathrm{cyc}}} \sum_{j=i-N_{\mathrm{half}}}^{i+N_{\mathrm{half}}} P(j,k)
\end{equation}

where $N_{\mathrm{cyc}} = 80$ samples/cycle and $N_{\mathrm{half}} = 40$. This introduces $\approx 8.3\,$ms look-ahead while maintaining sub-cycle latency.

The confidence is defined as $c(i) = \max_k P_{\text{avg}}[i,k]$, with predicted class $\hat{k}(i) = \arg\max_k P_{\text{avg}}[i,k]$. A decision is then emitted as

\begin{equation}
y(i) =
\begin{cases}
\hat{k}(i), & c(i) \ge \tau\\
-1, & \text{otherwise (abstain)}
\end{cases}
\end{equation}

where $\tau=0.6$ is a tunable threshold. Abstention prevents low-confidence misclassifications and, if sustained, flags the event as ``unclassified''—a safeguard for safety-critical grid operations. The complete online inference and decision process is summarized in Algorithm~\ref{alg:streaming}.

\begin{algorithm}
\caption{Streaming Inference with Centered One-Cycle Smoothing}
\label{alg:streaming}
\begin{algorithmic}[1]
\Require cycle length $N_{\mathrm{cyc}}{=}80$, half-window $N_{\mathrm{half}}{=}40$, threshold $\tau{=}0.60$
\State Initialize ring buffer $\mathcal{B}$ of length $N_{\mathrm{cyc}}$
\For{each new sample $i$}
    \State $\mathbf{p}\leftarrow \textsc{ModelPredictProba}(\mathbf{x}[i])$
    \State Append $\mathbf{p}$ to $\mathcal{B}$
    \If{$|\mathcal{B}|\geq N_{\mathrm{cyc}}$}
        \State Compute $\mathbf{q} = \frac{1}{N_{\mathrm{cyc}}}\sum_{j=i-N_{\mathrm{half}}}^{i+N_{\mathrm{half}}}\mathbf{P}[j]$
        \State $\text{conf}\leftarrow \max_k q_k$, \quad $\hat{k}\leftarrow \arg\max_k q_k$
        \If{$\text{conf}\ge \tau$} \State Emit $\hat{k}$ 
        \Else \State Emit $-1$ 
        \EndIf
    \EndIf
\EndFor
\end{algorithmic}
\end{algorithm}
\subsection{Latency-Aware Evaluation Metrics}

Models were evaluated on continuous time-domain signals to assess both correctness and responsiveness under real-time constraints. Four key metrics are defined:

\vspace{5pt}
\noindent\textbf{Accuracy.}  
The proportion of correctly classified samples:
\begin{equation}
\text{Accuracy} = \frac{\sum_i \mathbf{1}\{y(i)=y_{\text{true}}(i)\}}{N}
\end{equation}
where $y(i)$ is the predicted label, $y_{\text{true}}(i)$ the ground truth, and $N$ the total number of samples.

\vspace{5pt}
\noindent\textbf{Coverage.}  
The fraction of samples with a confident (non-abstained) decision:
\begin{equation}
\text{Coverage} = \frac{\sum_i \mathbf{1}\{y(i)\neq -1\}}{N}
\end{equation}

\vspace{5pt}
\noindent\textbf{Classification Time.}  
The classification time is defined as the interval between anomaly onset and the correct identification of its class:
\begin{equation}
T_{\text{cls}} = \min\{t : y(t) = y_{\text{true}}(t)\}
\end{equation}
This metric reflects the responsiveness of the model in assigning the correct label after an event occurs.

\vspace{5pt}
\noindent\textbf{Inference Latency.}  
Inference latency refers strictly to the computational delay of generating a decision from raw measurements. The per-sample forward-pass time is
\begin{equation}
T_{\text{inf}} = \mathbb{E}[t_{\text{exec}}]
\end{equation}
where $t_{\text{exec}}$ is the execution time of one forward pass, averaged over repeated runs after warm-up. $T_{\text{inf}}$ thus represents the average compute time required for a single inference.

\vspace{5pt}
\noindent These metrics capture correctness (\emph{accuracy}), decision availability (\emph{coverage}), responsiveness (\emph{classification time}), and deployability (\emph{end-to-end inference latency}) of the DL models.

\section{Results and Discussion}

\subsection{Training Performance Analysis}

Table~\ref{tab:training_results} summarizes the training results of eight deep learning models.  
All achieved strong performance with accuracies above 94\%.  
MLP was the top performer (99.66\% accuracy) despite its small size (49k parameters), showing that simple feedforward models can be highly effective.  
LSTM and CNN-LSTM surpassed 97\%, while Transformer, though less accurate (95.37\%), reached the highest balanced accuracy (99.88\%), demonstrating robustness to class imbalance.  
ResNet1D delivered solid accuracy but required nearly 1M parameters, highlighting the trade-off between performance and computational cost.

\begin{table}[!t]
\centering
\small
\caption{Model Training Performance Comparison}
\label{tab:training_results}
\begin{tabular}{@{}lrrrr@{}}
\toprule
\textbf{Model} & \textbf{Accuracy} & \textbf{Balanced} & \textbf{Macro F1} & \textbf{Parameters} \\
& \textbf{(\%)} & \textbf{Acc. (\%)} & \textbf{(\%)} & \\
\midrule
MLP & 99.66 & 99.88 & 99.53 & 157,394 \\
1D\_CNN & 98.89 & 99.77 & 98.71 & 285,842 \\
CNN\_LSTM & 97.13 & 99.65 & 96.84 & 542,290 \\
LSTM & 97.09 & 99.64 & 96.79 & 279,826 \\
ResNet1D & 96.21 & 99.52 & 95.87 & 977,106 \\
ST\_GCN & 95.89 & 99.49 & 95.52 & 67,730 \\
Transformer & 95.37 & 99.88 & 95.00 & 36,434 \\
TCN & 94.32 & 99.35 & 93.93 & 195,136 \\
\bottomrule
\end{tabular}
\end{table}

\subsection{Real-Time Streaming Inference and Latency Analysis}

Two representative streaming events were designed to evaluate anomaly detection and classification under realistic operating conditions with high-fidelity time-domain datasets. 
\textbf{Prediction Dataset~1 (Event~1)} consists of five consecutive anomalies, while 
\textbf{Prediction Dataset~2 (Event~2)} includes four anomalies. The complete ground truth is provided in Table~\ref{tab:ground_truth}.

\begin{table}[ht]
\centering
\caption{Ground Truth for Streaming Events}
\label{tab:ground_truth}
\begin{tabular}{l p{3.2cm} p{3.2cm}}
\toprule
\textbf{Time (s)} & \textbf{Event Stream 1 (Class)} & \textbf{Event Stream 2 (Class) } \\
\midrule
1.0–1.2 & SLG A-N (1) & ABC (14) \\
2.0–2.2 & LL B-C (7) & ABC-G (15) \\
3.0–3.2 & DLG AC-N (10) & - \\
4.0–4.2 & CT Attack MU32 (4) & GPS Attack MU32 (16) \\
5.0–5.2 & PT Attack MU23 (13) & A-G Fault in Zone 1 (-1) \\
\bottomrule
\end{tabular}
\end{table}

A series of streaming experiments was conducted using the two prediction datasets described above:
\vspace{5pt}
\subsubsection{Experiment 1: Streaming Predictions With and Without Cyclic Filtering}

\begin{figure*}[!t]
    \centering
    \vspace{-10pt}
    \subfloat[Raw streaming prediction (Event 1)\label{fig:raw}]{%
        \includegraphics[width=0.50\textwidth]{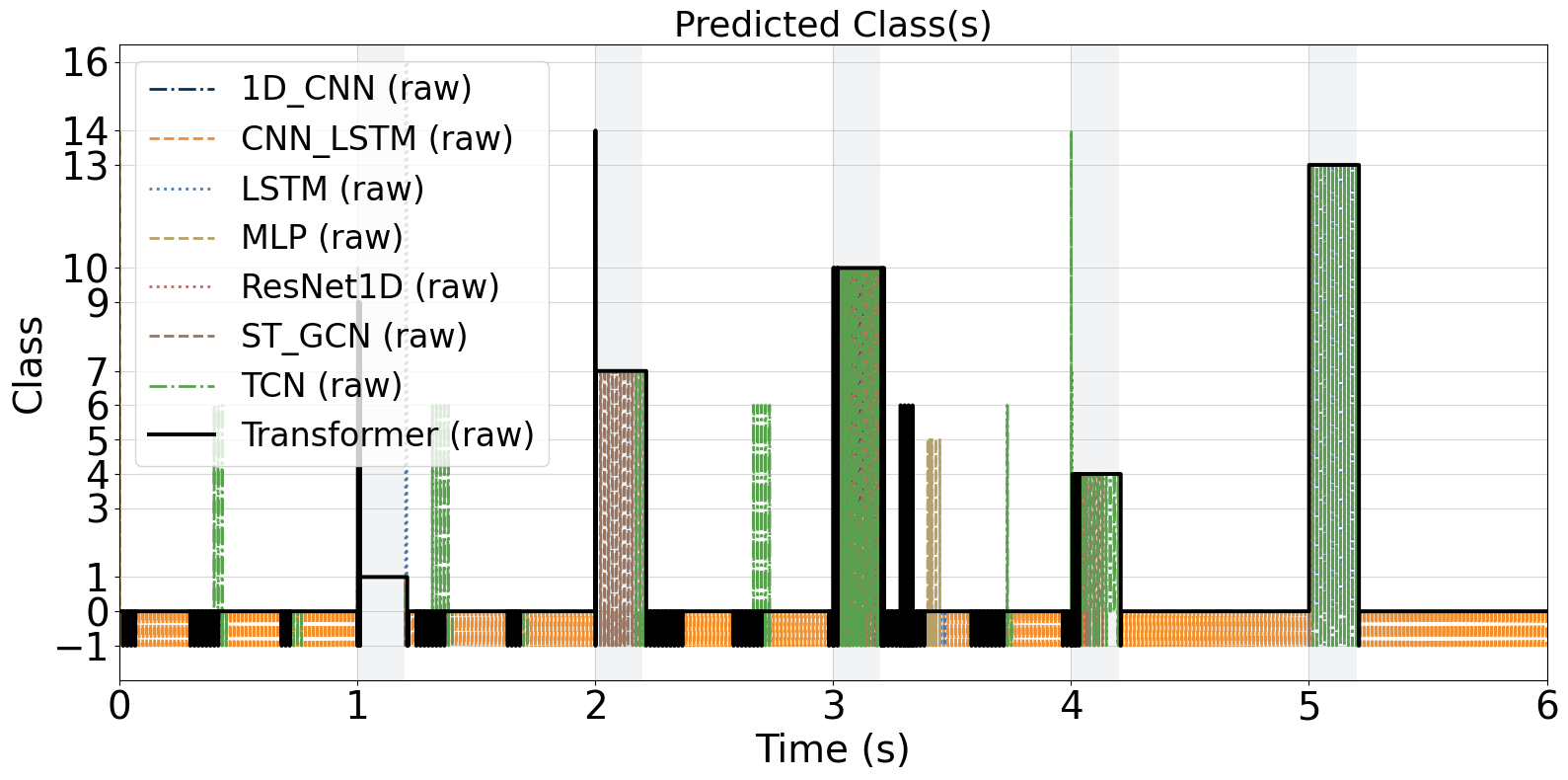}%
    }\hfill
    \subfloat[Prediction with one-cycle confidence averaging (Event 1)\label{fig:filtered}]{%
        \includegraphics[width=0.50\textwidth]{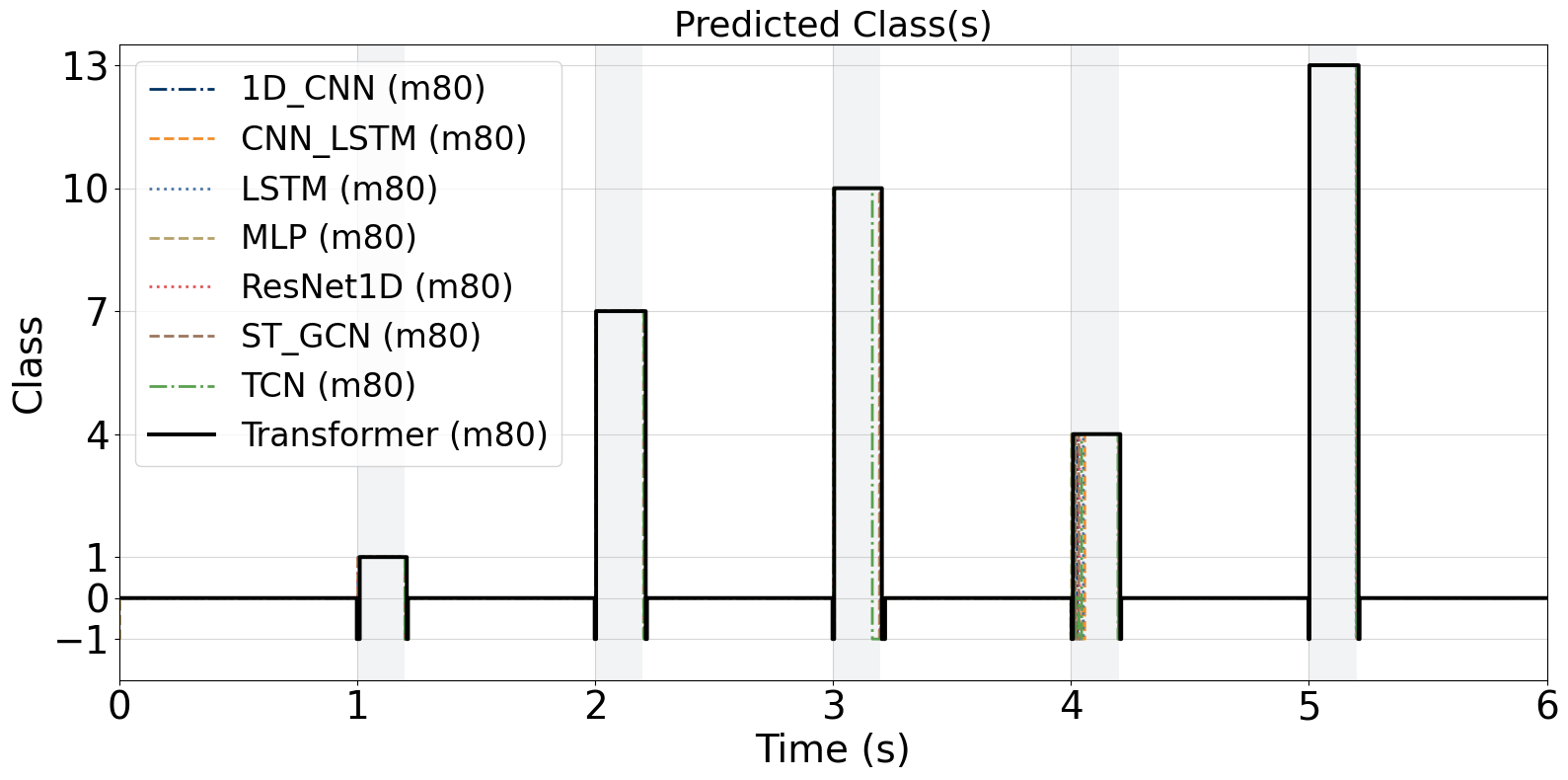}%
    }
    
    \caption{Streaming model outputs with and without cyclic confidence filtering. 
    Raw predictions (a) show high jitter and unstable class switching, whereas one-cycle moving average with confidence gating (b) yields smooth and decisive classifications with an explicit model confidence score.}
    \label{fig:model_responses}
\end{figure*}

The first experiment evaluated the stability of streaming classifications with raw, unfiltered data. As shown in Fig.~\ref{fig:raw}, the outputs exhibited substantial jitter and rapid oscillations between classes. This unstable behavior would trigger false alarms and is unsuitable for practical protection.

\vspace{5pt}
To address this, a one-cycle cyclic moving average filter with confidence gating was applied. As shown in Fig.~\ref{fig:filtered}, this resulted in classifications that were markedly more stable. The system was consistently identified as normal between anomalies, and abstentions were absent, reflecting sustained model confidence.

\vspace{5pt}

\subsubsection{Experiment 2: CT Ratio Attack }
The CT Ratio Attack on MU32 (Event 1, Period~4, Class~4) represents one of the most challenging anomalies, as it subtly scales current measurements rather than producing a clear disturbance.

\begin{figure}[h]
    \centering
    \includegraphics[width=0.9\linewidth]{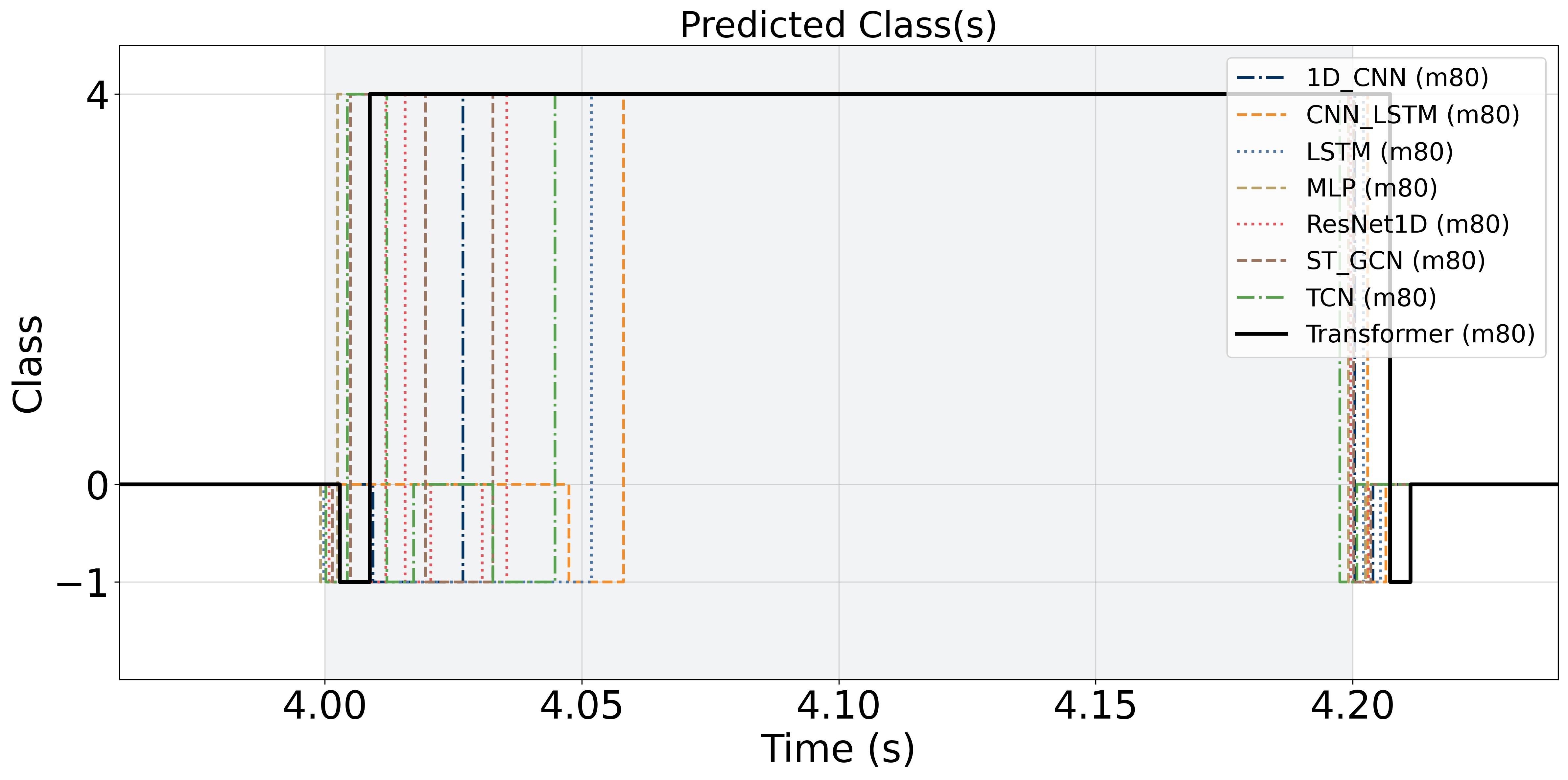}
    \caption{Zoomed-in predictions during the CT Ratio Attack.  
   }
    \label{fig:ct_attack_zoom}
\end{figure}

\vspace{5pt}
As shown in Fig.~\ref{fig:ct_attack_zoom}, the MLP and Transformer models reacted fastest and most reliably, achieving high coverage and correctness with classification times below 10~ms. In contrast, recurrent models such as LSTM and CNN--LSTM exhibited slower responses (above 50~ms) and lower coverage.

\vspace{5pt}
\subsubsection{Experiment 3: Multi-Event Classification}
\noindent The DL classification system was tested on Event 1 dataset containing five sequential anomalies. Fig.~\ref{fig:overview} illustrates the full sequence, where the model successfully tracks transitions between normal states, line faults, and cyberattacks. The model confidence dips align with the onset of each disturbance. 
Fig.~\ref{fig:ll_fault} zooms on a line-to-line (B–C) fault in Zone~2, showing accurate classification within milliseconds and stable confidence above the decision threshold.
\vspace{5pt}
\subsubsection{Experiment 4: Event 2 Cyberattack and Out-of-Zone Fault}
\begin{figure*}[!t]
    \centering
    \vspace{-5pt}
    \subfloat[Event 1: Overview of all anomalies (5 anomalies total)\label{fig:overview}]{%
        \includegraphics[width=0.48\textwidth]{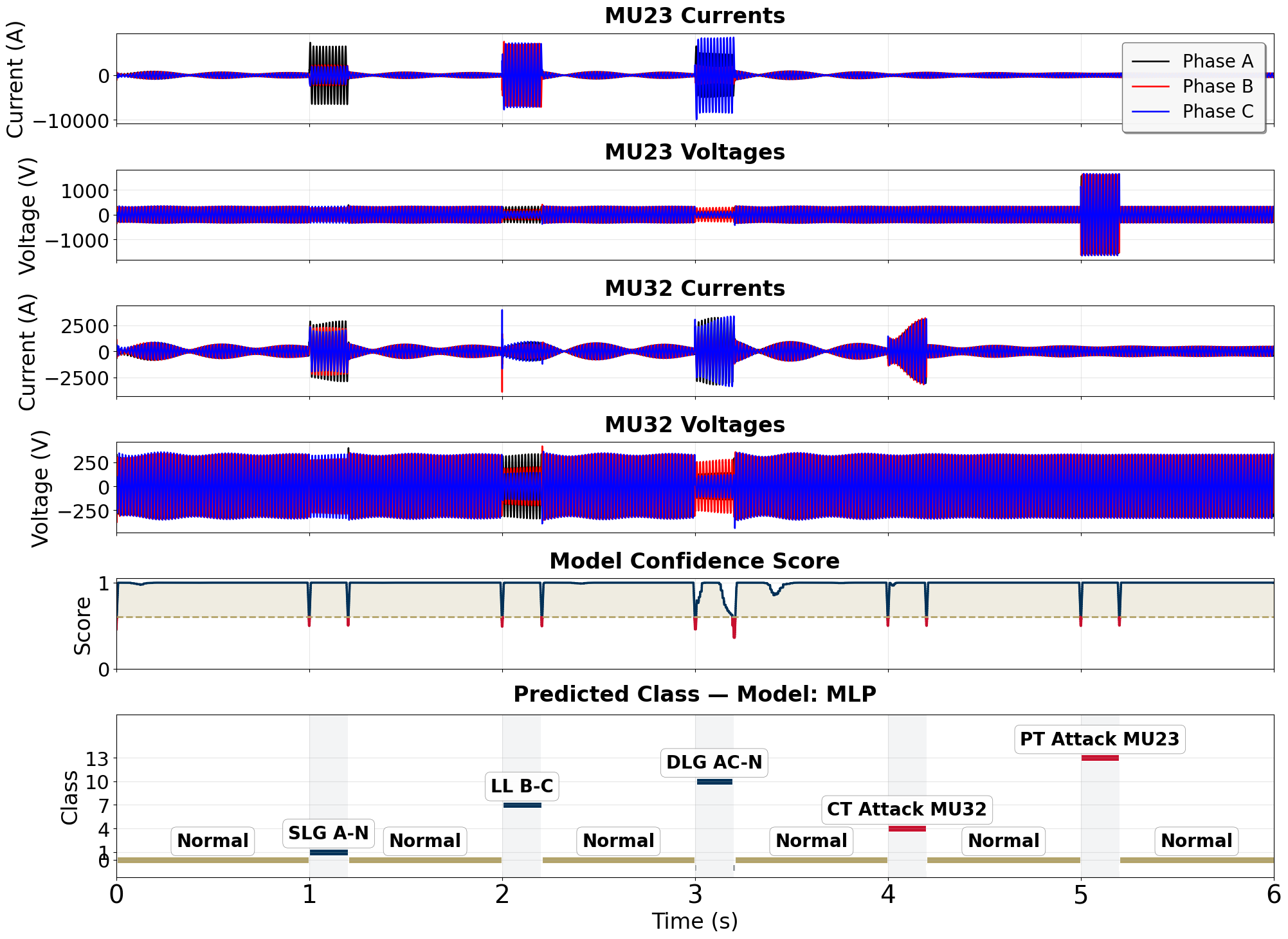}%
    }\hfill
    \subfloat[Event 1: Line-to-line fault (phases B–C, Zone 2)\label{fig:ll_fault}]{%
        \includegraphics[width=0.48\textwidth]{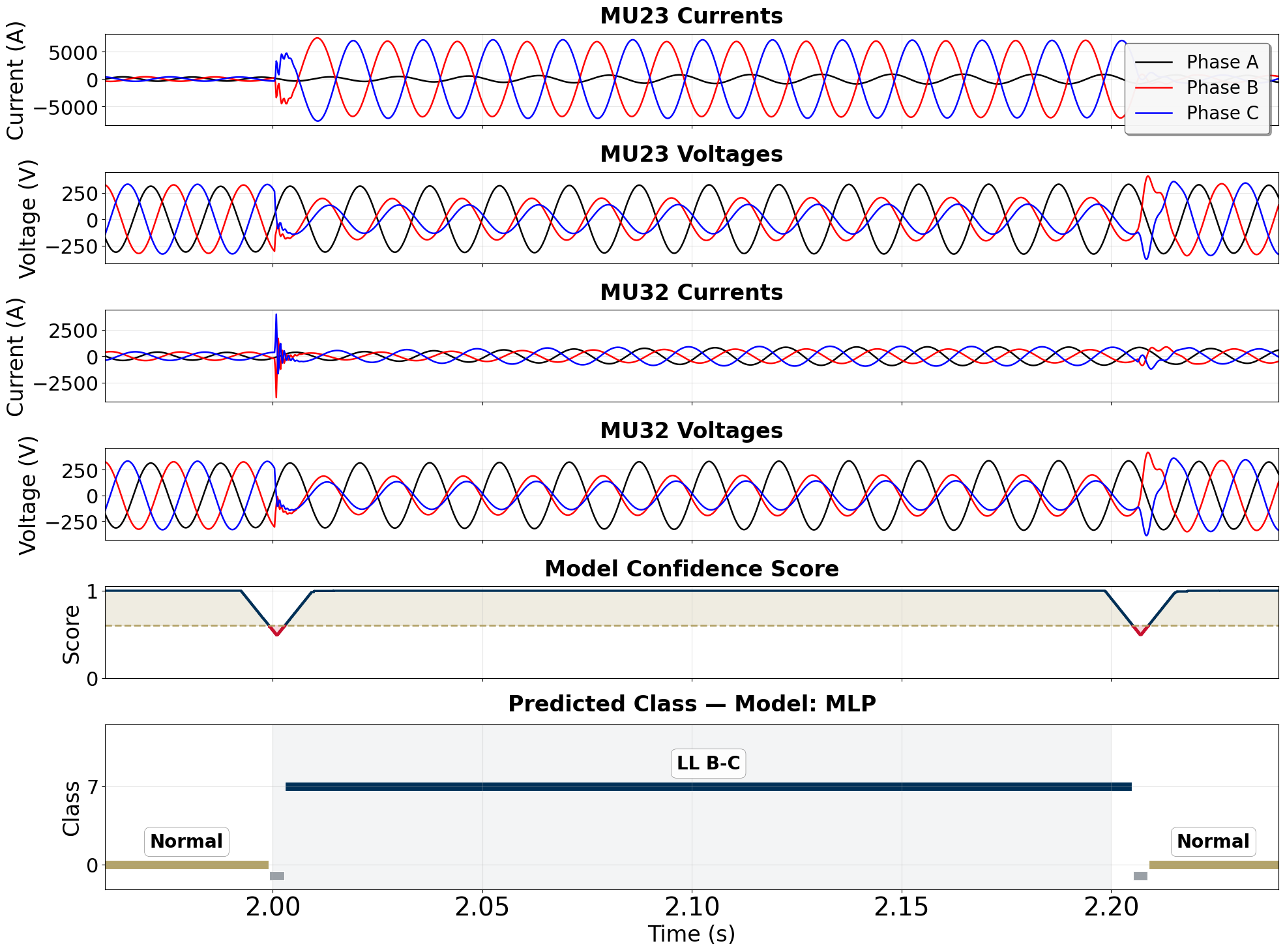}%
    }\\[1ex]
    
    \subfloat[Event 2: GPS spoofing at MU32 (4.0–4.2 s)\label{fig:gps_attack}]{%
        \includegraphics[width=0.48\textwidth]{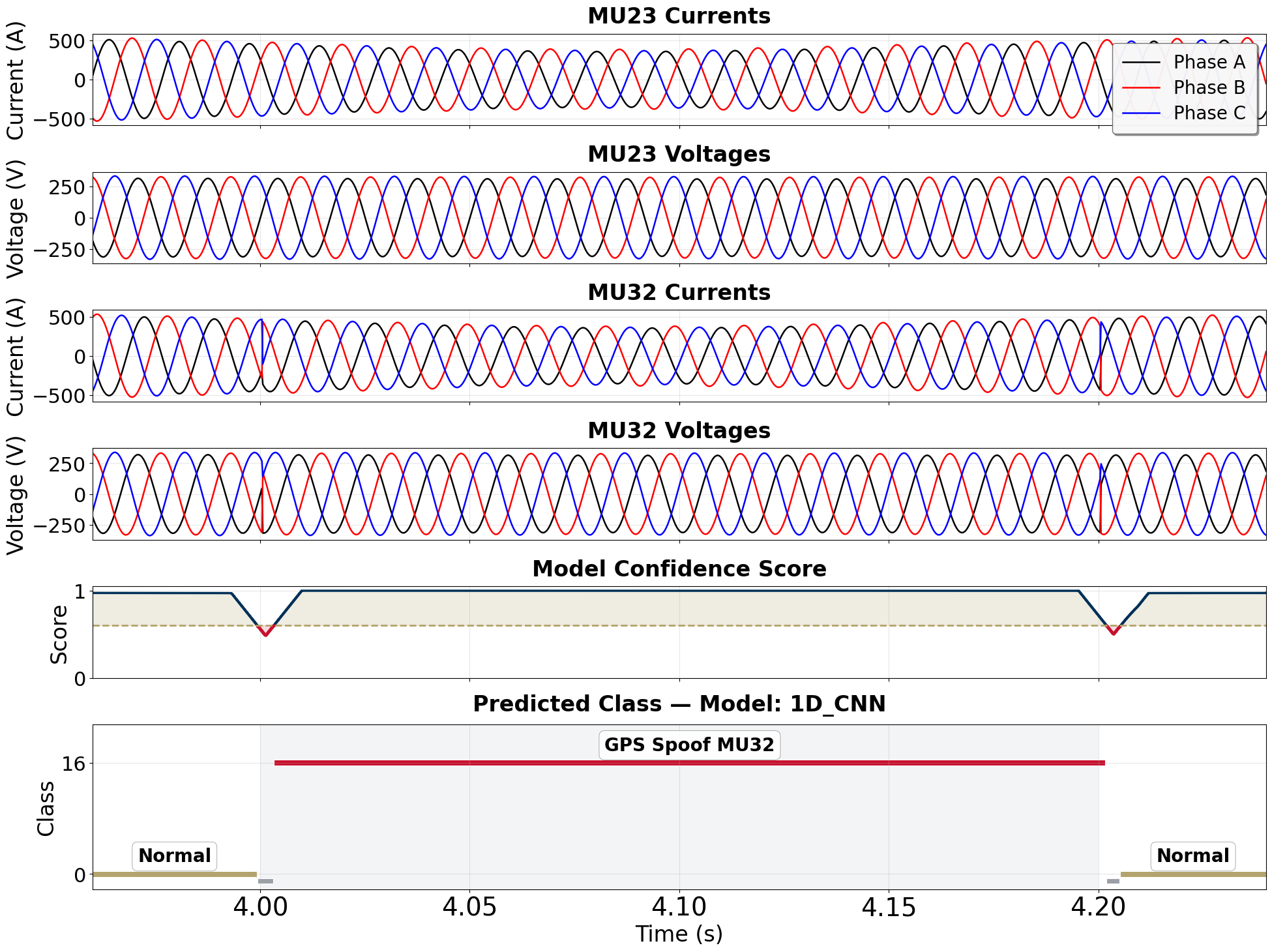}%
    }\hfill
    \subfloat[Event 2: SLG fault outside protection zone (5.0–5.2 s, classified as unseen $-1$)\label{fig:zone1_fault}]{%
        \includegraphics[width=0.48\textwidth]{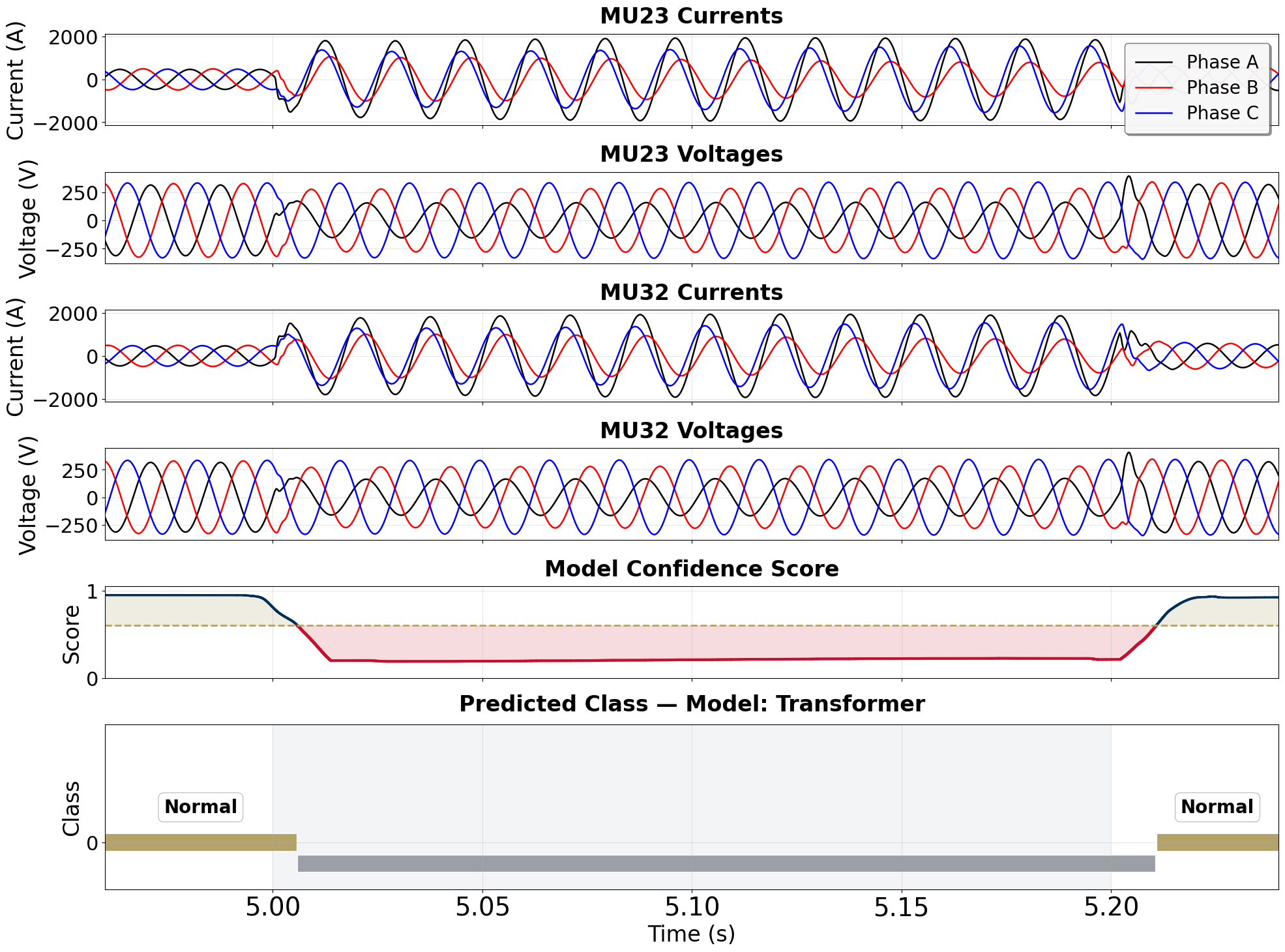}%
    }
    
    \caption{Model prediction analysis across two datasets of streaming events. 
    (a) Event~1 shows all five anomalies in sequence, (b) zoom on a line-to-line fault between phases B and C in Zone~2, (c) Event~2 highlighting GPS spoofing at MU32 between 4.0–4.2\,s, and (d) Event~2 showing an out-of-zone SLG fault (5.0–5.2\,s) that is detected by confidence drop and labeled as unseen class $-1$ since it was not part of training.}
    \label{fig:event_analysis}
\end{figure*}

\noindent The system was evaluated on two distinct disturbances.  Fig.~\ref{fig:gps_attack} highlights a GPS spoofing event at MU32 (4.0–4.2\,s), where the model promptly detects and labels the attack. 
Fig.~\ref{fig:zone1_fault} shows an SLG fault outside the training zone (5.0–5.2\,s). The classifier does not mislabel it as a known class; instead, it lowers its confidence, flagging the anomaly as an unseen class ($-1$).

\vspace{5pt}
\subsubsection{Experiment 5: Comprehensive Model Performance}
Each model was subjected to 100 runs of \textbf{Event~1} in order to evaluate average classification time, accuracy, coverage, and mean inference latency. 
The summarized results are reported in Table~\ref{tab:comprehensive}. The results confirm that classification time was consistently sub-cycle (3–15\,ms), which satisfies the fast response required for protection functions. However, the end-to-end mean inference latency exceeded three cycles (50–90\,ms), indicating that further optimization is necessary before these models can be deployed for primary protection. \\

At present, the latency profile is more suited for control and situational awareness functions, while future work should focus on deployment on high-performance industrial hardware to ensure compliance with protection-grade timing requirements. Limitations remain in the need for model retraining across different loading scenarios, and the inclusion of additional edge cases such as conductor loss and more diverse fault conditions. 
Addressing these factors is essential for ensuring robustness and generalizability in real-world deployment.

\vspace{5pt}
For reproducibility and comparison studies, all tests were executed on Google Colab (Linux kernel~6.1.123; 1 physical CPU core / 2 logical threads at~2.20\,GHz; 12.67\,GB RAM; single NVIDIA Tesla~T4 GPU; TensorFlow~2.19.0).\\

\begin{table}[t]
\centering
\caption{Comprehensive Model Performance Comparison}
\label{tab:comprehensive}
\small
\begin{tabular}{@{} l *{4}{c} @{}}
\toprule
\textbf{Model} & \textbf{Accuracy} & \textbf{Avg.} & \textbf{Avg. Class} & \textbf{Mean} \\
& \textbf{(\%)} & \textbf{Coverage} & \textbf{Time} & \textbf{Latency} \\
& & \textbf{(\%)} & \textbf{(ms)} & \textbf{(ms)} \\
\midrule
1D\_CNN & 95.73 & 96.67 & 8.54 & 70.22 \\
CNN\_LSTM & 92.59 & 97.34 & 14.83 & 69.64 \\
LSTM & 93.40 & 89.41 & 12.63 & 58.25 \\
MLP & 97.57 & 97.57 & 3.38 & 57.20 \\
ResNet1D & 95.03 & 96.11 & 5.29 & 72.42 \\
ST\_GCN & 96.44 & 96.59 & 4.25 & 88.89 \\
TCN & 90.88 & 92.50 & 4.17 & 61.68 \\
Transformer & 96.86 & 97.15 & 6.29 & 67.77 \\
\bottomrule
\end{tabular}
\end{table}
\section{Conclusion}

A latency-aware comparative study of eight deep learning architectures demonstrated that all models successfully classified physical faults and cyber-attacks using high-fidelity, time-domain waveforms from an industry-grade simulator. Across two streaming events, classification performance was uniformly accurate with sub-cycle decision times, confirming the feasibility of machine learning-based protection decisions at the waveform level. Despite this capability, inference latencies consistently exceeded three cycles, creating a critical limitation for primary protection deployment. The results highlight both the promise and the current limitations of deep learning for protection applications, pointing to the need for further optimization and hardware acceleration to close the latency gap and achieve operational readiness.

\bibliographystyle{IEEEtran}

\bibliography{references}

\end{document}